# Getting Users Smart Quick about Security

Results from 90 Minutes of Using a Persuasive Toolkit for
Facilitating Information Security Problem Solving by Non-Professionals

*Martin Ruskov, Paul Ekblom, M. Angela Sasse*
*University College London*

*Abstract*—There is a conflict between the need for security compliance by users and the fact that commonly they cannot afford to dedicate much of their time and energy to that security. A balanced level of user engagement in security is difficult to achieve due to difference of priorities between the business perspective and the security perspective. We sought to find a way to engage users minimally, yet efficiently, so that they would both improve their security awareness and provide necessary feedback for improvement purposes to security designers. We have developed a persuasive software toolkit to engage users in structured discussions about security vulnerabilities in their company and potential interventions addressing these. In the toolkit we have adapted and integrated an established framework from conventional crime prevention. In the research reported here we examine how non-professionals perceived security problems through a short-term use of the toolkit. We present perceptions from a pilot lab study in which randomly recruited participants had to analyze a crafted insider threat problem using the toolkit. Results demonstrate that study participants were able to successfully identify causes, propose interventions and engage in providing feedback on proposed interventions. Subsequent interviews show that participants have developed greater awareness of information security issues and the framework to address these, which in a real setting would lead ultimately to significant benefits for the organization. These results indicate that when well-structured such short-term engagement is sufficient for users to meaningfully take part in complex security discussions and develop in-depth understanding of theoretical principles of security.

I. Introduction

Users not complying with policies is a major concern in information security. There is growing evidence that a major part of non-compliance is due to misalignment between security and productive tasks. This is a consequence of both security designers not getting sufficient feedback from business users on how their solutions impact the core business process, and employees not being fully aware of the contribution of security to the business [1]–[3]. One possible way to improve this alignment is by involving business users in security discussions. For one thing, discussions might help improve mutual understanding and trust within the organization, for another they could help employees develop awareness about security problems and appreciation of the business value that security contributes.

Considering that employees are typically focused on business productivity, rather than security, their level of involvement in security discussions needs to be acceptable to them [3], [4]. Yet it also needs to help them to develop better awareness of their role in security. In turn this should mobilize them to avoid unintentionally exacerbating security problems e.g. by inattention to vulnerabilities. Finally, hopefully it would even lead them to actively prevent such problems through their daily behavior.

Such a change in perception becomes especially important when considering one of the key classes of threats in information security – those generated by insiders. Industry-wide surveys show that between a third and a quarter of information security attacks in organizations are performed by insiders [5]. When measured by resulting cost of attacks, these end up being as costly as outsider attacks. In some cases inside attackers are disgruntled employees seeking revenge on the company [5] who only partially appreciate the full risks to which they expose the organization [3]. In other cases employees are just being careless and unintentionally provide the opportunity for those carrying out the attack [3].

We set out to find a way to engage business users in security discussions in a way that is acceptable to them. One way to do this is to try to provide them with accessible and easy to use tools that would demand only short-term engagement. Ideally, despite such limited involvement, these tools will also lead users to develop improved understanding of the problem, and thus shape their behavior in future – hence the 'get smart quick' of the title. Part of our work on an early version of such a toolkit is presented here. Elsewhere we have already reported on the development process [6] and learning value [7] of this toolkit. Here we focus on user participation and perceptions, leaving examination of the value of collected feedback to security designers for further research.

In this paper we report the design, level of involvement and user perceptions of the security awareness toolkit that we have developed. Our development is based on an established crime prevention framework to guide users through a process they are unfamiliar with. The aim of the toolkit is to boost the security behavior of non-security employees, by engaging them in a relatively short discussion of the risks that their behavior might expose the organization to. In doing this we focus on users of secure systems employed by organizations and want to engage them to working towards being active *preventers* of security problems and not inadvertent *promoters*.

In the following sections we outline previous research that led us to this work, describe our crime prevention framework of choice, the procedure that we used to teach it and the toolkit that we developed for the purpose. At the end we summarize the results of our analysis of participant involvement and self-reports and the conclusions that these results have led us to.

## II. BACKGROUND

Enabling discussions about perceived inadequacies of security measures has been widely acknowledged as a key element to improving security in organizations [3], [5], [8]. This has led various researchers to employ technology to educate non-security staff, see e.g. [9], [10]. In several studies on factors contributing to information security compliance, [11], [12] it has been confirmed that employees' *threat appraisal* (i.e. appreciation of the actual severity of risks) does affect attitude towards complying.

Other research [5] has looked into understanding and early detection of insider attacks. Yet, there is a need for more discussion of the interpersonal and social factors that enable insider threats, be it through the absence of *crime preventers* (people who by their action or presence makes crime less likely), or the presence of collaborators (or more generally speaking *crime promoters*). Taking a broader perspective, it becomes apparent that these environmental factors do interact with the personal and technical ones already studied. Some approaches, e.g. the one taken by Blackwell [13], consider the interplay of different factors, but not the social environment. Sasse and colleagues explore the notion of *security culture* [1], [4], but do not discuss in depth the role that individuals play in shaping it. Similarly in Siponen, Pahnila and Mahmood's work *social influence* is represented to the extent it impacts normative beliefs. They verify the role of these beliefs into forming employees' intention to comply [11]. In their more recent study [12] the authors mention that "social influence works both ways", thus alluding to the possibility that promoters can turn into preventers or vice versa, but they do not actually explore the idea further beyond the suggestion.

Another related field of research – conventional crime prevention – has accumulated more experience, in this regard, than information security. Research here, under the discipline of crime science has included social aspects when considering both personal and situational factors that play a role in crime. While most approaches in crime science have focused on the crime situation, more holistic perspectives which integrate this with a richer understanding of the offender have been claimed to be advantageous [14], [15].

However, the consideration of these resulting combinations leads to an increase of complexity which, although necessary for getting to grips with real-world crime, makes the problematic considerably harder to grasp. This is even more challenging when the framework needs to be brought to the attention of employees who are not professional security staff, thus not having security as their primary task. Yet, if people get more efficient in understanding the role of security, they will see the need for and understand the value of applying it [3], [4].

Again, a solution to a similar challenge has been found in other research disciplines. The fields of persuasive technology and experiential learning have been exploring how technology can be utilized to support comprehension and awareness, by fostering both engagement and learning. This challenge actually corresponds to a particular factor contributing towards information security compliance behavior, as identified by Pahnila and colleagues [11]. This factor is *information quality* as perceived by employees and in their study it was shown to influence actual information security compliance.

One pervasive approach in employing technology to help overcome complexity is what BJ Fogg has called *persuasive technology* [16] – utilizing technologies to cause attitude or behaviour change. In his book on the topic, he identifies three functional uses of technology for the purposes of persuasion: as a *tool*, as a *medium*, and as a *social actor*.

Fogg explicates that technology as a tool can *simplify*, *tunnel* (i.e. guide through a process) and *adapt* to the individual's needs. His interpretation of technologies as a medium concerns their use to simulate and elicit relationships between causes and effects. Finally, in order to deliver technologies as social actors they need to be attractive and provide a feeling of *similarity* to the people using them. Technology needs to offer *praise*, demand *reciprocity* from users and provide a feeling of *authority*. Some of these persuasive techniques have been used by researchers at Carleton University to design usable security tools aimed at non-security staff [9].

In his book Fogg [16] acknowledges the relatedness of his concept of persuasive technology to techniques used in games. When compared to serious games or gamification, the concept of persuasive technologies is more formal, but relatively new and consequently its use for development has been studied less. A number of platforms claim to employ game-based approaches, including variations of the persuasive technology techniques reviewed here, for the purposes of comprehension and awareness development. However, reviews [17], [18] show that most of them still fall short of verifying whether they help their users grasp the knowledge that they claim to convey. This is a problem that has been widely discussed in relation to learning technologies and Laurillard [19] has proposed a framework to compare the taught object being presented to the understood object actually being comprehended. In that *conversational framework*, she suggests that this process of transferring understanding goes through the constructed learning environment and the actual actions that learners do to engage with it. As a consequence, the analysis of how learning happens needs to consider both of these as well. Such an analysis would provide evidence of the information quality of the toolkit.

## III. THE CCO FRAMEWORK

As an instantiation of a holistic approach that considers technical, personal and social factors, we have turned to the

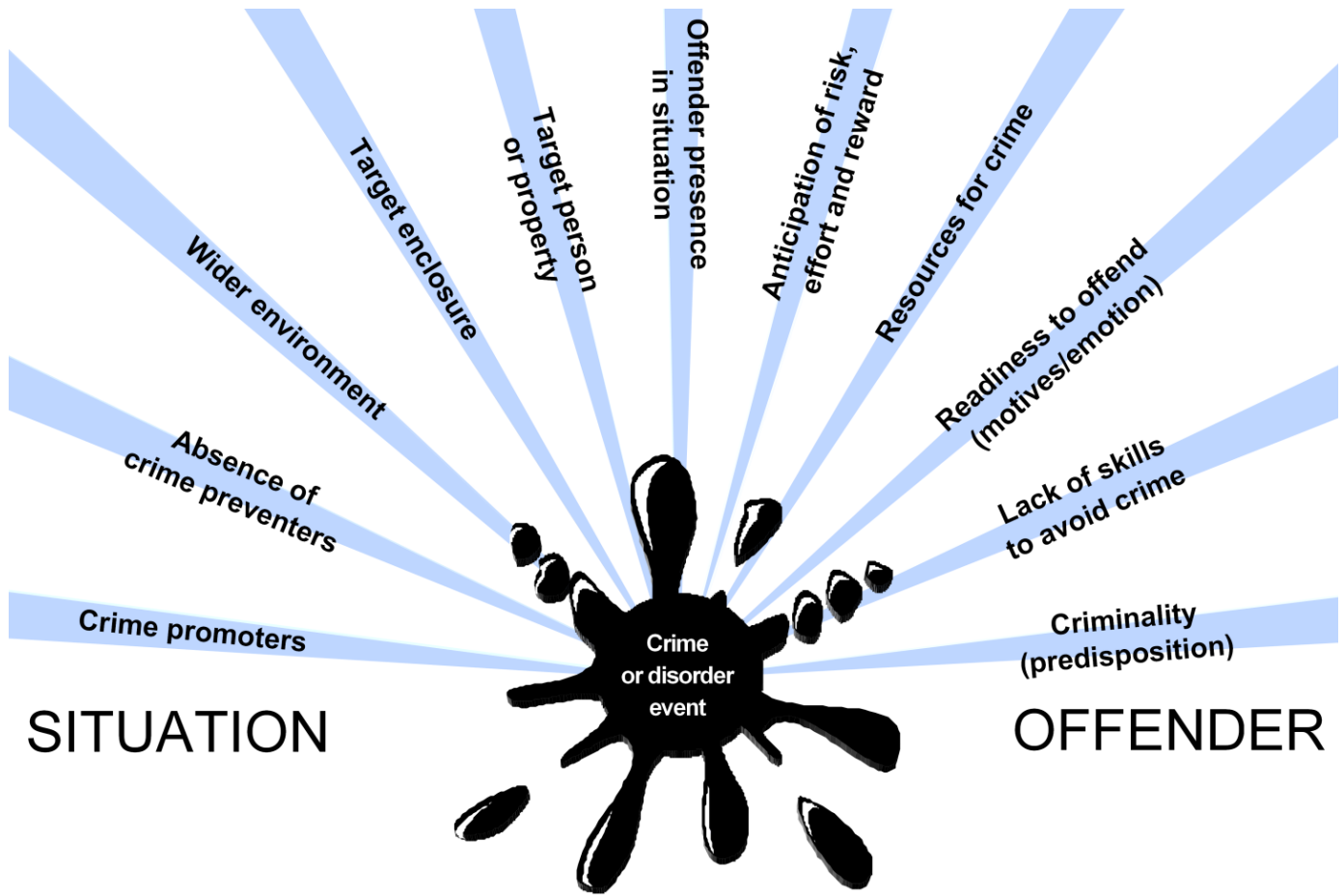

Fig. 1. The Conjunction of Criminal Opportunity (CCO) diagram, as presented in [21]. It features the eleven generic causes that, when combined, give rise to the criminal event. On the left-hand side of the diagram are represented situational factors, and on the right-hand side are factors brought in from the perspective of the potential attacker (i.e. offender).

Conjunction of Criminal Opportunity framework [15]. This framework, CCO for short, combines offender-oriented and situational factors that contribute to a conjunction of immediate causes allowing a criminal event to occur. The necessary preconditions for this to happen are as follows:

- An offender who is *predisposed* and *ready* to commit crime
- Perceiving an acceptable risk of *harm, effort/cost/time, reward*
- Properly-equipped with *tools* and *perpetrator techniques*…
- Encounters a valuable and insecure *target* (property or person)
- In the absence of people who can act as ready, willing and able *preventers* (e.g. professional security officers, or any employee or other individual who by their presence or action makes the crime less likely)
- In the presence of people who can act as *crime promoters* (someone who inadvertently, carelessly or deliberately makes the crime more likely to happen, e.g. by leaving their desktop computer unattended and unlocked)
- In an *environment* and perhaps an *enclosure* (e.g. secure office space or VPN) which contains attractive targets, which routinely brings offenders and targets together and whose properties favor offender over preventer.

The circumstances enabling the conjunction that leads to the criminal event listed above have been visualized on a diagram (see figure 1) representing them as rays coming together into a blot – the criminal event itself.

The CCO framework was developed as a generalization resulting from the analysis of several thousand of crime prevention projects implemented through the UK Safer Cities program [20]. Its typical original application was in the analysis of criminal hotspots wherein certain types of crime recur in a small vicinity.

CCO was designed to contribute to the *preventive process*, a very basic outline of which is set out in the following steps [21]:

- Identification of crime problem – the symptoms – and setting of objectives for reduction

- Diagnosis of causes of crime problem
- Selection of specific interventions, and creation of practical operational solutions
- Implementation
- Evaluation and adjustment

A more advanced equivalent of this process is in [20]. Considering each of the eleven generic causes in the CCO diagram, naturally leads to ideas for their intervention counterparts. These intervention ideas could block, weaken or divert the causes, such that the criminal event is less likely to be attempted, or to succeed. For example an attempt to reduce the absence of crime preventers could be the introduction of security staff to undertake surveillance; an example for securing the enclosure could be the introduction of access control; and the anticipated risk, effort and reward could be reduced by means of deterrence like the introduction of penalties.

The variety of possible intervention ideas and the exact details of their implementation lead to a classification of how specific, or general these ideas are. The CCO framework distinguishes between *principles* and *methods*. Methods are the detailed practical activities that are necessary to be conducted to implement the intervention. Research shows that these are often difficult to transfer to other situations – what works in crime prevention is very context-dependent [22]. Principles, on the other hand, are the more abstract description of what is being done that is formulated in a way that could be re-applied, customized to context in other situations. Let's consider the possibility of employee satisfaction probes. The principle behind them could be the need to check, and then if necessary address, employee attitudes for disgruntlement (as an instance of '*readiness to offend*'). A common implementation method involves collecting satisfaction feedback at annual appraisal interviews, and of course acting on the results.

This example already shows that there is more than one method of implementing an intervention. In this particular case an alternative could be providing an anonymized feedback box. On the other hand the method of annual interviewing can also be used to implement other intervention principles: involving staff in discussions about potential process improvements could be interpreted, say, as an instance of the broader principle of mobilizing people as crime preventers. From these examples it should be apparent that many-to-many relationships are possible between intervention principles and intervention methods.

## IV. STUDY METHOD

This paper reports some qualitative outcomes from a wider experimental study to develop and evaluate an interactive toolkit using persuasive techniques to teach elements of security thinking to non-specialists in the domain of insider threats. The work is at an exploratory phase, so the experimental conditions are only reported in brief. To introduce the study, this section reports how participants were recruited, what procedure was employed and how data was analyzed.

### A. Participants

To recruit participants for this study we put an advertisement in a university recruitment pool and offered a financial reward. Subsequently 28 of the applicants were recruited to participate in the experiment.

19 male and 9 female participants took part. Their age ranged from 20 to 65 with an average of 26.5 and median 23.5. Five of those participants reported that they had some previous exposure to information security or a related field. This was in general only limited and ranged from training on information security in the army or at university, to deploying firewalls and anti-virus software. None of them had actual professional experience.

### B. Procedure

For the purposes of the experiment participants were split into control and experimental conditions. When invited to the experiment by e-mail, participants were given two one-page-long texts to digest. One of these was an introduction to the CCO framework and the other was a problem scenario, describing an insider threat and set out in detail below. Both of these texts were available as reference to participants throughout the study.

The laboratory session lasted 90 minutes and involved using the toolkit for up to 60 minutes with two tests (before and after). Both tests aimed to assess participants' knowledge of crime prevention and the CCO framework in particular, in the course of i) analyzing the causes of the exemplar crime problem; and ii) coming up with, then assessing, preventive solutions. When using the toolkit participants were deliberately not given instructions, unless they asked for them. If they didn't suggest any ideas, they were told they needed to come up with at least two to continue. In general they were not reminded of time, unless they took too long in the first idea generation steps (refer to the explanation of the toolkit for the full process in the materials section below). If this happened, the researcher present in the room told them that there was more to do in the toolkit and that they should finish with the idea they were currently writing down and proceed with subsequent steps. Time was planned so that they finished at least 10 minutes before the full 90 minutes were over, to give them time to do the final test. After that test participants were interviewed about their experience. They were also invited to take part in a subsequent competition, in which they had the chance to use the toolkit further and the ones that ranked first (see below) were offered a cash prize. After the end of this competition further interviews with a sample of the participants were conducted. Both interview sessions were recorded and the researcher took notes.

In this paper only experiential and qualitative results are reported. The experiential part of the results are recorded contributions from users during the working through of the toolkit. The qualitative part includes analysis of the two interview sessions – at the lab session and the subsequent later interview.

### C. Analysis

As already mentioned the wider study was a controlled experiment. The difference between control and experimental conditions was that participants in the former were required to assess the ideas they had previously generated *themselves* on causes and intervention, whereas those in the latter were shown ideas of *someone else* to assess. This is further clarified in the section dedicated to the toolkit in the materials section below. The distinction between the two conditions is of minimal relevance to this paper – for present purposes the results reported for both conditions were equally valid and hence combined here.

Thematic analysis was used to code the interview data. This was done based on the researcher's notes. The codes were subsequently expanded with partial transcriptions from the interview recordings. Finally, following Laurillard's conversational framework, the interviews were interpreted to compare participants' understanding against the original CCO framework.

When reporting quotes from the interviews codes were used to anonymize participants. The codes would typically consist of a letter indicating the study group (E for experimental and C for control), a double-digit participant number and a letter indicating when the quote was made (I for immediate, and D for delayed). A typical example for a code would be C04D, indicating that the participant assessed their own ideas with the toolkit, and that the quote is from their second interview.

## V. MATERIALS

The materials that were necessary for this study were all embedded in a bespoke website. These were an introduction to the CCO framework, the tailor-made problem scenario and the actual web-based toolkit used to navigate participants through the framework, to provide the necessary guidance and feedback and to actually collect the user-generated data. The materials described here are accessible online at http://cco.ko64eto.com.

### A. The CCO Introductory Text

CCO is a multifaceted and complex framework. For the purposes of the study we had to come up with an introduction that would be short enough to fit on one page – to ensure it did not take up too much time within the study sessions or dissuade participants. Faced with this challenge, we decided to focus on information that would not be easily interpretable from the toolkit itself, while at the same time providing the necessary background knowledge and awareness to allow for a quick start.

The eleven casual elements of CCO can be read off the diagram when participants use it. The distinction between principles and methods, on the other hand, is better understood when grounded in examples. That's why we decided to provide a functional description, rather than a factual one. It was based on the preventive process, as it gave an overarching view of how causes and interventions (the objects of interaction) fit in the bigger picture. This description did not include the eleven elements, nor did it include the diagram, details about the use of the diagram, or explanation of intervention principles and methods. As it will be seen in the following dedicated section, these were left to the toolkit to take care of.

### B. Problem Scenario

In order to give participants the opportunity to practice application of the framework, the scenario had to introduce them to a realistic description of a relevant recurring problem. On one hand this had to be a situation without a seemingly straightforward solution that would have left them with the feeling that they had solved it in just a few minutes. On the other, its emphasis had to allow them to identify employees as a factor in solving it.

The scenario that we developed described recurring insider attacks and was based on survey data from The CERT Guide to Insider Threats [5]. As a way to make it representative we designed it to apply to two of the three most commonly recurring sources of insider threat: IT sabotage and theft of intellectual property.

The scenario describes a frequently relocated outsourcing center for IT services where disgruntlement among staff grows to the extent that rapid turnover leads to a hit-and-run culture of insiders attempting to make a big win at the company's expense. The text featured a sample of six cases of insider attacks, aimed to illustrate their diversity. These ranged from numerous activations of virus protection software to leaking sensitive data or poaching customers when leaving the company.

### C. Web-based Toolkit

The toolkit guided participants through a process consisting of four consecutive parts: introduction to the scenario, idea generation, idea assessment and score review. This was the instantiation of Fogg's *tunneling*. The toolkit is an instance of using persuasive technology as a medium in that it helps participants understand cause-and-effect relationships. Examples are provided later in this section.

The scenario part essentially provided participants with the opportunity to re-read the scenario already encountered in the introduction to the session. At the end of this step, as in general with the process, they were able to determine for themselves when they were ready to move on to the idea generation part. This was meant as a form of personalization by making the toolkit adapt to participants' own pace.

The idea generation part of the toolkit featured an interactive version of the CCO diagram. It is provided in two consecutive modes – identification of *causes* and suggestion of *interventions* (the latter being shown in figure 2). Each of those allowed participants to focus on one single ray of the diagram by clicking on it. When they did that on the causes diagram, a dialogue box appeared, that provided an explanation and an example for that particular causal ray. In this dialogue box participants were provided with the opportunity to write down suggestions of how this generic cause is being instantiated in the scenario at hand, which is an instance of how the toolkit simplified participants' work with the CCO framework. Similarly, in the interventions diagram a dialogue box provided context to participants by reminding them of their suggested causes, and typical intervention principles relevant to the corresponding generic cause. In the background the toolkit employed a simple word-based pattern-matching algorithm [23] to try to match newly suggested ideas to existing ones. When the algorithm couldn't match a new contribution to any existing with sufficient certainty, the new idea was annotated as being innovative and a "new" icon popped up on the participant's screen. This icon was intended as a form of immediate *praise* to participants who come up with new ideas.

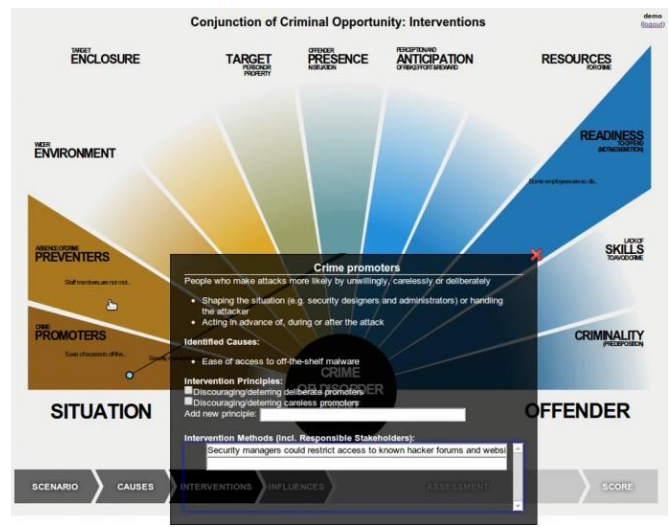

Fig. 1. The idea generation screen, as recreated by the toolkit for the identification of interventions. It features visual feedback for identified ideas and semi-structured input dialogue for participants to suggest ideas. This and the following screenshots are provided for illustrative purposes and the contained texts are not essential to this publication.

After these phases of generation participants were given access to a table having the proposed interventions as rows, and the eleven causes as columns (see figure 3). This way they could further explore the influence a suggested intervention could have on wider causes and correspondingly how it is interconnected with other interventions.

Both the set of causes and interventions diagrams on one hand, and the influences table were examples of exploring causes and effects using persuasive technology as a medium.

The assessment part of the toolkit prompted participants to assess ideas of intervention methods. This is where the experimental and control groups differed. As said, the control group got to assess their own method ideas, whereas the experimental group assessed a predetermined set of ideas, meant to be seen as someone else's contribution. Participants were asked to grade each idea along a 5-point Likert-scale and were provided with an empty text field if they wanted to provide further comments to clarify their assessment.

Each participant made three distinct assessments. As a way to encourage participants to explore the problem from different perspectives, each of these assessment was presented through the eyes of a different role in the scenario. The first one was from the perspective of the security designers – the role of the *professional crime preventer* that participants assumed so far in their engagement with the toolkit. The second assessment (see figure 4) was from the perspective of the insider (the *offender* in CCO terms) assessing how ideas impacted their perception of the criminal opportunity. In the last assessment players were put in the role of corporate management (key *neutral citizen*) who could potentially become *victim* of crime, helpful *preventer* or even unintentional crime *promoter*.

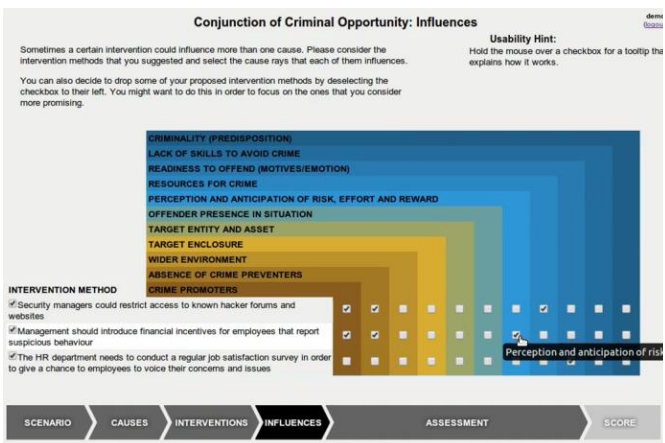

Fig. 3. The toolkit screen that asks participants to think whether their proposed interventions could influence other causes, beyond the one that they originally designed it for.

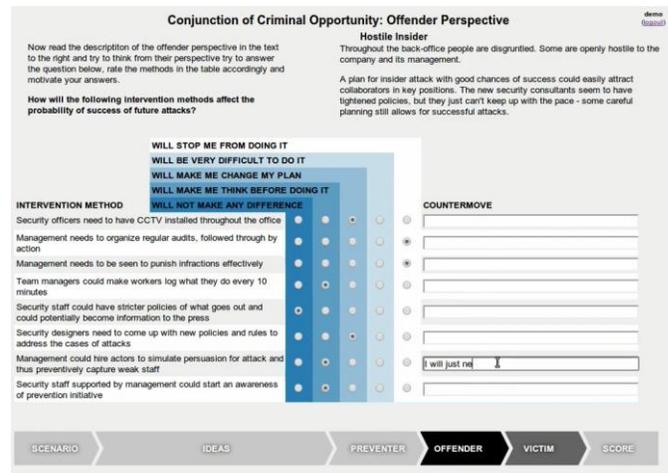

Fig. 4. The second of the three assessment screens. Here participants are asked to assess how interventions will affect success estimations by offenders.

There was also variation in the questions posed to participants. In the first and second assessments they rated how the implementation of each intervention idea would affect the probability of further attacks. In the third they assessed how far intervention ideas may have reduced the harm that the criminal event might cause. In effect, these three perspectives addressed risk as a breakdown into probability (the first two assessment) and harm (the last assessment).

These assessments, as well as the score and ranking screen that is to be explained subsequently, enabled the toolkit to provide participants with a sense of *reciprocity*.

The final toolkit part was a score and ranking screen (figure 5) that showed how that particular participant performed. This included a table with intervention ideas, suggested by the participant, and overall statistics and ranking of their performance. The table featured a breakdown of the three scores these ideas cumulatively received from other participants.

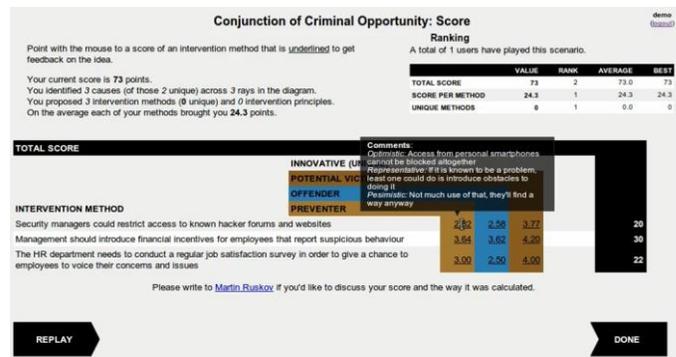

Fig. 5. The final score screen. In the upper part of the screen are score and ranking. When a participant hovered with the mouse over an individual assessment of an idea, a pop-up with the comments given to this assessment appears.

This last screen also summarized these scores into participants' overall score and ranking. Participants saw their provisional ranking in the experimental group according to three distinct metrics: 1) overall score; 2) average score per intervention method; and 3) number of proposed innovative method ideas (i.e. ones that the pattern-matching algorithm couldn't identify with existing ones).

Delivering assessment and then seeing how this reflected in the final score was also an example of using persuasive technologies as a medium.

## VI. RESULTS

There are two types of study results reported here: quantitative summary of ideas suggested by participants within the toolkit; and quotes from the interviews conducted after the study to elicit participants' perceptions of the experience. The first part shows the extent to which participants got engaged with the toolkit; the second, their perception of the toolkit and the CCO framework.

### A. Involvement with the Toolkit

The participants in this study were able to effectively identify causes and propose interventions, even though they were not security experts. Figure 6 shows the number of ideas and comments that participants generated during their use of the system. On average participants came up with 11 causes (typically one per CCO cause ray), 8 intervention methods and two comments. Typically people that had the opportunity to comment on ideas of others came up with more comments than those that commented on their own ideas.

### B. 5.2 Perceptions of the CCO Framework

The study reconfirmed that the CCO framework on its own is difficult to grasp, because of its complexity and the effort needed to apply a general framework to a specific context. One participant suggested a formulation of what she identified as a possible obstacle. In the interview she phrased it as an open questions: "*how to actually analyze the data in a form which can be useful... a crime is not a math thing, which you can analyze, it's a big and complex thing. So all I thought is how you actually identify and find that useful information which can help you prevent the next time...*" (C06I)

Participants also made more focused comments about the particular aspects that were difficult. Many spoke of the ambiguity of the eleven generic causes. A few participants were critical of it, but there were two that appreciated the ambiguity in depth. One of them said the diagram "*looks a bit daunting. All these words you look at them and wonder what they mean. Some points are either too similar – I don't like that, but I can see the need for it. Or they seem too relevant, but I am not sure if I am giving the right information. It seems a bit overwhelming. I can see the need for it – if you need ten things, I have repeated some things because of that. The explanations are OK, if they weren't there I'd wanted them*" (E02D). The other commented on the balance of the number of rays talking that there is "*a lot of overlap, but not so much that any of these is redundant*" (E05D).

Many other participants went on to provide examples and recommendations of what didn't make sense to them and how the diagram could be rearranged. For the purposes of this paper it suffices to note that these comments are indications of the fact that after the use of the toolkit they were able to critically reflect on the CCO framework.

Two other themes that participants discussed were the difficulty of instantiating principles into new ideas and the challenge when participants had to think of the interconnectedness of those ideas. One of them said that "*brainstorming, new ideas – these were difficult*" (C14I), another one that they "*had to think about causes, effects, suggestions, ideas... this was very active. You had to do a lot of thinking and formulating for a technical subject*" (C25D). The interconnectedness of ideas was approached in the causal influences screen. That prompted one participant to comment that influences were "*too tedious*" (E27D). When talking of that screen another one said to have "*found [it] to be quite complex. And I guess it is complex because of all the interconnecting ideas... The layout itself is complex, because the ideas are complex...*" (C15D).

### C. Misconceptions of the CCO Framework

There were a number of participants that explained difficulties in understanding that they still had despite having used the toolkit.

Talking of the framework one participant for example explained that the "*model applies to virtual… does not apply enough to a real world*". Then she elaborated: "*in the online world the model makes perfect sense. In terms of actual reality people are more complicated. Why they do things, sometimes you need to go back to their childhood. Online stuff is all about information, presentation, transparency, the public persona of a company*" (C15I).

Another participant explained that she found explanations and terminology difficult to understand. Her further explanations showed her confusion: "*What I found difficult with other questions that followed was when it said that you have to come up with theories. So for me everything that was practical about the scenario and every time I had to give a practical solution or a practical explanation I found that easy because I could easily put myself in the scenario and imagine how things would work out. But then when I was trying to create theories, and trying to come up with abstract things again, then I wasn't really sure what to do.*" (E18D)

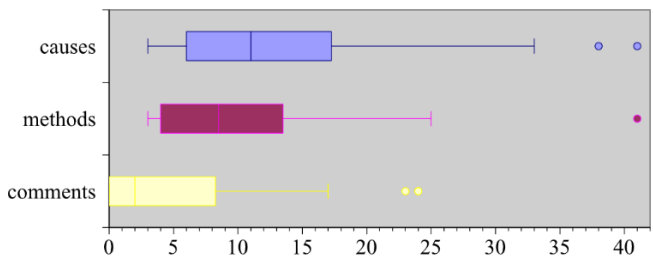

Fig. 6. A box plot with outliers showing the number of user contributions when working through the toolkit.

*D. Perceptions of the Toolkit*

In this section we report participant comments that illustrate how they perceived the experience. These show that participants were attracted to use the toolkit, they were engaged while using it and they found something for them personally in it. All this goes to show that they did feel the beneficial effect of persuasive technology applied to the CCO framework.

During the interviews several participants showed appreciation of the relevance of the scenario – something that arguably helped them engage with it. One participant said that it felt as a "*very live issue. And there's that companies are moving back and forth all over the world. This is a real time situation. So it is a very practical thing and that's why I took it so seriously*" (C04D). Another one explained that he was "*not necessarily looking at a scenario where you have a development company or development team, developing software, but in an everyday organization, which is IT-centric, so they are using IT to take up their business. They are going to have an IT department. There is still opportunity for this kind of crime in any company*" (C01D). A third participant commented that it is an "*interesting case, because it shows how morale and employee treatment are intangible causes that lead to such tangible effects. That was a very interesting case... It sounds exaggerated, but I could foresee such cases happening probably to a lesser extent*" (E08D). Yet another participant spoke of the value she saw in the examples: "*It was interesting, because it is big, important subject. You know, for personal users and businesses as well. So it did open my mind to all of the different forms of attack on computer systems*" (C25D).

Despite the widely acknowledged complexity of the framework among participants, there was also recognition of its being made accessible by the toolkit. In the words of one participant they gradually developed their understanding. He said that it was "*a lot of information coming at once. At first it is a bit overwhelming and after a while I got the gist of it.*" (C26I)

Two main themes that could be identified are that the toolkit shaped the interaction of participants and that it helped them focus on particular aspects of the problem while maintaining coherence of the big picture. Regarding shaping interactions, one participant explained that he "*did it one by one. I looked at the different causes and then I went back to the scenario and looked what could have applied to that cause... of course [the causes] helped. Because it gets you to look at the scenario from a lot of different angles. Because you look at each cause, and then you go back to the scenario and look at it with that cause in mind*" (E05D). Another referred to the specific form this information was provided in, saying that "*it's good that [suggestions are in] a pop-up because once you finish this you don't need to see it again because you will be moving to another part.*" (C25D). The last example also sets the other relevant theme here – letting participants focus on just parts of the solution at a time. Just after she had done that, one participant noticed that in her words "*I repeated a few of them... I didn't realize that I am repeating it, but when I saw them all together I thought that they have become irrelevant because I have already thought of them once before*" (C07I). All in all, one participant summarized that she found the task "*interesting, because it is clear and good. If you click on it there is clear explanation... [it] has a lot of information. It can be used by non-professionals*" (E19I).

Different participants explained what got them involved. Some participants spoke of the challenge for them to understand how the toolkit (and inherently the CCO framework) works. One of them said that she "*found the mechanics behind it interesting, the process of it*" (C15D). Another participant valued the challenge, saying: "*The enjoyment, it was interesting, seems very real. How can you prevent that, to deal with the employees? I was attracted also from a professional point of view.*" (E02D). One participant even discussed the aim of the toolkit: "*if the objective of the [toolkit] was to really make you think in detail about what things can affect the security of the place. Maybe it is good in the sense that it makes you think in a well-rounded view*" (C26D).

Participants also acknowledged that once they got involved with the toolkit, it also kept them engaged beyond just completing the task. One said: "*I did a lot of repetition. I was doing the task again, like being back here. I did it at home*" (E02D). Another explained his feelings about the opportunity to do so: "*I like the idea that I could get back home and then do it again, so there isn't much time pressure involved and I can do it in my own free time*" (E27D). Another one actually spoke of getting engaged in the topic beyond the toolkit "*I did a bit of research, had some friends in [the domain], checked out websites. It was something I always want to know*" (C04D).

A few participants shared that they felt encouraged by some persuasive elements. Talking of the new idea icon on the idea generation screens one participant explained "*Yes, and it was quite encouraging when the new idea [icon] came up. Because I'm not really good with this kind of things. So when I write something and it says "new" then I was quite encouraged and I would think of other points that would give me the new idea thing*" (E09D). Another one explained how competing involved her, elaborating on some of the related factors: "*I liked that there are three different [score] categories. So it gives a good idea of what exactly can you be the top in... If I know which is the best score I could try to work harder on it... The problem was there are some contestants who do the final boost. So towards maybe the last one hour they will type in all the ideas and overtake the first [competitor]*" (E27D).

Several participants acknowledged that the toolkit allowed them to have a distinctive personal take of the framework. This had to do with both personal interpretation of the scenario and choosing how exactly to use the toolkit.

In line with the first observation, several participants explained that their contributions were based on their personal experience. One explained it in her own words: "*my answers depended a lot on my background. I am working towards being a corporate lawyer. A couple of my answers are based on this background*" (E27D). Another found the scenario difficult and explained that "*I just knew it was something about IT there... [my ideas were based on] my own feeling and also the examples given in the rays*" (E09D). One participant elaborated that she found only some rays to be relevant: "*I was thinking about all of [the causes] and for me from the scenario there were just a few causes that could have created the problem. And I don't think

*that all those causes that the computer was offering were necessarily there in the scenario*" (E18I).

There were similar reflections explaining selective involvement in providing comments in the assessment screens: "*when I rate [an idea] lowly I would try to give my comment to explain why I thought it was bad, but if I ranked it highly then I wouldn't bother commenting*" (E27D).

*E. Frustration with the Toolkit*

While some participants were positive about the experience, others were less enthusiastic. In this section we report critical opinions expressed. Although there were many specific recommendations regarding the usability of the toolkit, these are not reported here, being beyond the scope of this paper. Instead the focus here is rather on complaints that are inherent to the approach.

One aspect that was perceived negatively was realism. To one participant (C04D) the scenario had "*too short a period*" and "*too much stuff happening.*" Another commented that in the scenario "*there's a lot of information. It's quite heavy*" (E02D).

A few participants (E05D), (E19D) commented that they didn't know that comments in the score were written by someone else. This might have undermined participants' appreciation of the fact that they are actually engaging in a constructive dialogue about ideas. One participant commented that some of the ideas they had to assess were "*not realistic*" (E27D), which might allude to the fact that the toolkit used pattern-matching against previous ideas to provide immediate feedback, and in rare cases this lead to inaccurate attributions of ideas to comments.

Some participants were confused about what they should do. One of them explained "*to a certain extent unless if you were here, I wouldn't have been able to comprehend what I am supposed to do... If you weren't here and I only have these instructions I am not really certain I would be able to completely grasp the concept.*" (C15D). Others were more specific saying that it was "*lengthy in terms of words.*" (E27D).

*F. Improved Awareness of Information Security and the Application of the CCO Framework*

Several participants explained their (presumably newly developed) understanding of the complexities involved in information security. In the words of one of them he had a chance to "*realize that crime is not only about opportunity. It goes beyond what you see in the dictionary*" (C15I). Others went into more detail. One acknowledged that "*if a security manager fails... he's a crime promoter effectively, maybe not deliberately...*" (C17I). A few participants from the experimental group (as they were the ones subjected to it) acknowledged the benefits of peer-learning, e.g. explaining that the study "*...involved a lot of thinking and analyzing techniques. I could see how some of the ideas other people have come up altogether later could be used to improve security methods and preventive techniques...*" (E24I).

Participants also elaborated on how they understood why CCO makes a distinction between intervention principles and intervention methods. One explained that "*intervention principles are more general, and intervention methods are specific methods. A principle can have different methods, but they can come from a single principle*" (C06I). Another one explained that using the toolkit they can think of the different causes that one method can address "*[in interventions screen] I can only fill them each separately. But [in influences screen] a single intervention can have several influences at the same time. So here it is more [focused] on that what the different influences are.*" (C06I).

VII. DISCUSSION

In about an hour the toolkit engaged each participant in a structured discussion about insider threat problems. In this session, they managed to come up with a range of ideas for causes and interventions.

Feedback collected during the two subsequent interview sessions reconfirms that the CCO framework is commonly perceived as too complex despite the guidance employed. This complexity confused several participants to the point that they misinterpreted what is its purpose. Still, generally participants managed to grasp the essence of the framework and understand how to use it. The fact that they found the eleven generic cause categories vague and overlapping, didn't stop most of them from actually working with them. Some of them appreciated that this had to do with the actual interplay across causes and methods. The collected evidence demonstrates that participants understood the importance of both social influence and threat appraisal. It also allows for a clearer interpretation of the specific opportunities to improve these further, esp. when engaging with the issues of a particular organization. One might speculate that if the problem scenario had more closely reflected the everyday experience of participants they might have found it easier to apply the CCO framework.

Furthermore, this toolkit could be adapted to the needs of organizations with bespoke scenarios that more closely correspond to the actual issues of the organization and would be more relevant to its employees. Doing this to get employees to discuss suggestions will both allow for a new source of ideas, but also for better awareness about current insider threats.

Participants showed evidence that the toolkit helped them to both understand and use the CCO framework. The interviews reconfirmed that the techniques borrowed from the persuasive technology literature were helpful. Participants acknowledged that the toolkit guided them to understand key aspects of the CCO framework better, but also saw it as challenging to work with. This effective persuasion demonstrated by the toolkit is an example of the application of security awareness techniques with high information quality.

Overall, for many participants this session was enough to provoke deeper interest in information security and to help them develop relevant interpretations of the framework. While these participants had no previous experience with the CCO framework the toolkit, despite being an early iteration with some clear shortcomings, nevertheless allowed them to use it effectively.

These results indicate that this approach of involving users in security discussions through persuasive technologies like the toolkit used here shows some early positive results. When given to users for a short period it generally does improve their understanding of possible information security risks to the company and how they personally could either promote or prevent them. In combination with other research [11] this gives rise to the expectation that this will lead users to consider how to reduce their own contribution to possible risks and even engage them in taking the initiative on more extended risk prevention, for example by reporting noticed vulnerabilities to security staff.

Findings so far give us confidence to continue and take the toolkit out of the lab and further develop it for use in class, and ultimately in an actual organization in the industry.

In the vein of applied empirical research our results indicate that when well-structured such short-term engagement is sufficient for users to meaningfully take part in complex security discussions and develop in-depth understanding of theoretical principles of security – in other words, to get smart, quick.